\documentstyle[pra,aps,multicol,epsfig]{revtex}

\input epsf.tex

\newcommand{\be}{\begin{equation}}
\newcommand{\ee}{\end{equation}}
\newcommand{\ba}{\begin{eqnarray}}
\newcommand{\ea}{\end{eqnarray}}
\newcommand{\ban}{\begin{eqnarray*}}
\newcommand{\ean}{\end{eqnarray*}}

\newcommand{\ket}[1]{\mbox{$ | #1 \rangle $}}

\newcommand{\demi}{\frac{1}{2}}

\newcommand{\myone}{\leavevmode\hbox{\small1\normalsize\kern-.33em1}}

\begin{document}

\title{Photon-Number-Splitting versus Cloning Attacks in Practical Implementations
of the Bennett-Brassard 1984 protocol for Quantum Cryptography}
\author{Armand Niederberger$^{1}$, Valerio Scarani$^{2}$, Nicolas Gisin$^{2}$}
\address{$^{1}$ Section de Physique, Ecole Polytechinque F\'ed\'erale de Lausanne,
CH-1015 Ecublens\\ $^{2}$ Group of Applied Physics, University of
Geneva, 20, rue de l'Ecole-de-M\'edecine, CH-1211 Geneva 4,
Switzerland}
\date{\today}
\maketitle

\begin{abstract}
In practical quantum cryptography, the source sometimes produces
multi-photon pulses, thus enabling the eavesdropper Eve to perform
the powerful photon-number-splitting (PNS) attack. Recently, it
was shown by Curty and L\"utkenhaus [Phys. Rev. A {\bf 69}, 042321
(2004)] that the PNS attack is not always the optimal attack when
two photons are present: if errors are present in the correlations
Alice-Bob and if Eve cannot modify Bob's detection efficiency, Eve
gains a larger amount of information using another attack based on
a $2\rightarrow 3$ cloning machine. In this work, we extend this
analysis to all distances Alice-Bob. We identify a new incoherent
$2\rightarrow 3$ cloning attack which performs better than those
described before. Using it, we confirm that, in the presence of
errors, Eve's better strategy uses $2\rightarrow 3$ cloning
attacks instead of the PNS. However, this improvement is very
small for the implementations of the Bennett-Brassard 1984 (BB84)
protocol. Thus, the existence of these new attacks is conceptually
interesting but basically does not change the value of the
security parameters of BB84. The main results are valid both for
Poissonian and sub-Poissonian sources.
\end{abstract}

\begin{multicols}{2}

\section{Introduction}

Quantum cryptography, or more precisely quantum key distribution
(QKD) is a physically secure method for the distribution of a
secret key between two distant partners, Alice and Bob, that share
a quantum channel and a classical authenticated channel
\cite{review}. Its security comes from the well-known fact that
the measurement of an unknown quantum state modifies the state
itself: thus an eavesdropper on the quantum channel, Eve, cannot
get information on the key without introducing errors in the
correlations between Alice and Bob. In equivalent terms, QKD is
secure because of the no-cloning theorem of quantum mechanics: Eve
cannot duplicate the signal and forward a perfect copy to Bob.

However, perfect single-photon sources are never available, and in
most practical implementation the source is simply an attenuated
laser. This means that some of the pulses travelling from Alice to
Bob contain more than one photon. These items, in the unavoidable
presence of losses in the quantum channel, open an important
loophole for security: Eve may perform the so-called
photon-number-splitting (PNS) attack, consisting in keeping one
photon in a quantum memory while forwarding the remaining ones to
Bob \cite{brassard,lutkenhaus}. This way, Eve has kept a perfect
copy without introducing any error. In particular, here we
consider the BB84 QKD protocol introduced by Bennett and Brassard
in 1984 \cite{bb84}. In this protocol, when the basis is revealed
in the sifting phase Eve can measure each photon that she has kept
in the good basis and obtain full information on the bit.

Until recently, it was thought that this attack was the best Eve
could do when two or more photons are present. However, in a
recent work \cite{cl}, Curty and L\"utkenhaus (CL) have shown that
this is not the case for noisy lines (optical visibility $V<1$)
and imperfect detectors (quantum efficiency $\eta$, dark count
probability $p_d$), when the natural assumption is made that Eve
cannot modify the detectors' parameters. Basically, the idea is
simple: consider pulses that contain two photons. In the PNS
attack, Eve has full information after the basis announcement {\em
provided} Bob has detected the photon that was sent. So, in the
information balance, Eve's information for such an item is
$\eta\times 1$. Suppose now that Eve, instead of performing the
PNS, uses a suitable $2\rightarrow 3$ cloning machine, keeps one
photon and forwards the other two photons to Bob. Eve's
information conditioned to Bob's detection could be $I_{c2}\leq
1$, but now the probability that Bob detects a photon of the pulse
is $(1-(1-\eta)^2)$. Thus for small values of $\eta$, Eve's
information for a two-photon pulse becomes $2\eta\times I_{c2}$,
and this may be larger than $\eta$. Of course, by using such a
cloner, Eve introduces some errors, so this attack is possible
only up to the expected quantum bit error rate (QBER).

As we prove below however, the analysis of CL is restricted to a
specific distance of the line Alice-Bob, which turns out to be
unrealistically short. The goal of this paper is to evaluate the
contribution of the individual attacks that use $2\rightarrow 3$
cloning machines for all distances in a realistic range of
parameters. When this is done, the contribution of attacks using
$2\rightarrow 3$ cloning machines leads to a negligible
improvement over the usual PNS strategies: both the achievable
secret-key rate and the maximal distance are for all practical
purpose the same, whether these new attacks are used or not. This
is our main result. In the run, we describe a new strategy that
uses a $2\rightarrow 3$ cloning machines, that performs better
than those previously described. This new strategy has an
intuitive explanation which opens the possibility of immediate
generalizations: in particular, it may prove useful to study the
security of other protocols, against which the PNS attacks are
less effective \cite{sarg,sarg2,decoy}.

The paper is constructed as follows. In Section \ref{sec2}, we
state precisely our hypotheses and write down general formulae, in
which Eve's attack is parametrized by the probabilities of
performing each strategy, and submitted to some constraints. At
the end of this Section, we show that the analysis of CL, correct
though it is, is valid only for a given distance between Alice and
Bob, whence the need for the present extension of their work.
Section \ref{sec3} contains the main results: we perform numerical
optimization assuming the two known $2\rightarrow 3$ cloning
strategies and our new one, showing that ours performs indeed
better but that its contribution is on the whole negligible.
Section \ref{secnew} is devoted to some extensions and remarks.
Finally, in Section \ref{sec4}, we give some semi-analytical
formulae that reproduce the full numerical optimization to a
satisfactory degree of accuracy: these are useful for
experimentalists, to find bounds for the performance of their
setups. Section \ref{sec5} is a conclusion.

\section{Hypotheses and general formulae}
\label{sec2}

\subsection{Imperfect source, line and detectors}

We are concerned with practical quantum cryptography, so the first
point is to describe the limitations on Alice's and Bob's
hardware. We work in a prepare-and-measure scheme.

{\em Alice's source.} Alice encodes her classical bits in light
pulses; the number of photons in each pulse is distributed
according to a probability law $p_A(n)$. In most practical QKD
setups, Alice's source is an attenuated laser pulse, so
$p_A(n)=p(n|\mu)$ the Poissonian distribution of mean photon
number $\mu$. But our general formulae and most of our results
will be valid independently of the distribution, so in particular
they apply to all quasi-single-photon sources \cite{single}. For
heralded single-photons obtained from an entangled pair
\cite{herald}, the situation is more complex. If the twin photon
is used only as a trigger, and the preparation of the state is
done directly on the photon(s) travelling to Bob, then this source
behaves exactly as a sub-Poissonian source, and our subsequent
analysis applies. If on the contrary the twin photon is used also
for the preparation (because one detects its polarization state,
thus preparing at a distance the state of the photon travelling to
Bob), then the PNS attack is not relevant
\cite{review,lutkenhaus}.

{\em Alice-Bob quantum channel.} The quantum channel which
connects Alice and Bob is characterized by the losses $\alpha$,
usually given in dB/km (for optical fibers at the telecom
wavelength 1550nm, the typical value is $\alpha\simeq 0.25$dB/km).
The transmission of the line at a distance $d$ is therefore \ba
t&=& 10^{-\alpha\,d/10}\,. \ea Moreover, we take into account
non-perfect visibility $V$ of the interference fringes.

{\em Bob's detector.} It has a limited quantum efficiency $\eta$
and a probability of dark count per gate $p_d$. The gate here
means that Bob knows when a pulse sent by Alice is supposed to
arrive, and opens his detector only at those times; so here, "per
[Bob's] gate" and "per [Alice's] pulse" are equivalent. Those two
parameters are not uncorrelated: in reverse-biased avalanche
photodiodes, a larger bias voltage increases both $\eta$ and
$p_d$. Typical values nowadays are $\eta=0.1$ and $p_d=10^{-5}$.

\subsection{Alice and Bob's rates and information}

We write $p_B(0)$ the probability per pulse that Bob detects no
photon sent by Alice. Since both losses in the line and detection
are binomial processes, \ba p_B(0)&=&\sum_n
p_A(n)\,(1-t\eta)^n\,;\ea for a Poissonian distribution on Alice's
side, $p_B(0)=p(0|\mu t\eta)$. We consider only those cases in
which Alice and Bob use the same basis, because in any case the
other items will be discarded during the sifting phase. Bob's
count rates per pulse in the "right" and the "wrong" detector are
then given by \cite{note5} \ba
C_{right}&=&\demi\,\left[\big(1-p_B(0)\big)\left(\frac{1+V}{2}
\right)\,+\, p_B(0)\,p_d\right] \\
C_{wrong}&=&\demi\,\left[\big(1-p_B(0)\big)\left(\frac{1-V}{2}
\right)\,+\, p_B(0)\,p_d\right]\,,\ea where the factor $\demi$
accounts for the losses in the sifting phase. The QBER is the
fraction of wrong bits accepted by Bob, \ba
Q&=&\frac{C_{wrong}}{C_{right}+C_{wrong}}\,=\,
\demi-\frac{V}{2\left(1+\frac{2p_dp_B(0)}{1-p_B(0)}\right)}\,. \ea
In particular, as long as $\big(1-p_B(0)\big)\gg p_B(0)\,p_d$, one
can neglect $C_{wrong}$ in the denominator and decompose
$Q=Q_{opt}+Q_{det}$, with the optical QBER defined as
$Q_{opt}=\frac{1-V}{2}$. The mutual information Alice-Bob after
sifting is \ba I(A:B)&=&\left(C_{right}+C_{wrong}\right) \,
\left[1-H(Q)\right] \label{iab}\ea where $H$ is Shannon entropy.

\subsection{Hypotheses on Eve's attacks}

{\em Hypothesis 1:} The characteristics of the quantum channel
(the optical QBER, or more precisely $V$, and the losses, that
determine the transmission $t$) are fully attributed to Eve. On
the contrary, Eve has no access to Bob's detector: $\eta$ and
$p_d$ are given parameters for both Bob and Eve. The eavesdropper
will of course adapt her strategy to the value of these
parameters, but she cannot play with them. This hypothesis is
almost unanimously accepted as reasonable; it implies that Bob
monitors the rate of double clicks when he happened to measure in
the wrong basis; if this rate is larger than expected, he aborts
the protocol. As realized by CL \cite{cl}, it is precisely this
hypothesis that opens the possibility for the cloning attacks to
perform better than the PNS \cite{note6}.

{\em Hypothesis 2:} Through her PNS attacks, Eve should not modify
Bob's expected count rate due to Alice's photons
$C_{ph}=\demi[1-p_B(0)]$. This constraint is usually assumed in
the study of PNS attacks, see e.g. Refs
\cite{brassard,lutkenhaus,cl,sarg,sarg2}; still, two comments are
needed. One could strengthen the constraint by requiring Eve to
reproduce the full photon-number statistics at Bob's side. But one
could as well weaken it: here, we are asking that Bob should not
notice PNS attacks at all; Eve could be allowed to perform
noticeable PNS attacks, in which case one should bound her
information and study the possibility of privacy amplification.

{\em Hypothesis 3:} Eve performs incoherent attacks: she attacks
each pulse individually, and measures her quantum systems just
after the sifting phase. The justification for this strong
hypothesis is related to the state-of-the-art of the research in
quantum cryptography: no one has found yet an {\em explicit}
coherent attacks that performs better than the incoherent ones
\cite{singap}. In other words, incoherent attacks are still used
to compute upper bounds for security, while "unconditional
security" proofs provide lower bounds \cite{uncond}, and for all
protocols there is an open gap between the two bounds. Note also
that incoherent attacks are not "realistic" in the sense of those
described e.g.~in \cite{real}; in particular, Eve is allowed to
store quantum information in a quantum memory. The hypothesis of
incoherent attacks implies in particular that after sifting,
Alice, Bob and Eve share several independent realizations of a
random variable distributed according to a classical probability
law. Under this assumption and the assumption of one-way error
correction and privacy amplification, the Csiszar-K\"orner bound
applies \cite{csi}: one can achieve a secret-key rate given by \ba
S&=&I(A:B)\,-\,I(A:E)\,. \label{rsk}\ea Actually, this is a
conservative assumption: in the presence of dark counts,
$I(B:E)<I(A:E)$ holds, so the strict bound for $S$ is
$I(A:B)-I(B:E)$; however, the difference is small, and $I(A:E)$ is
easier to estimate. We devote paragraph \ref{ssieb} below to
comment about $I(B:E)$. The mutual information $I(A:B)$ has been
given in (\ref{iab}), we should now provide an expression for
$I(A:E)$.

\subsection{Eve's strategies}
\label{sseve}

Having stated the hypotheses on Eve's attacks, we can now
formulate Eve's strategy as a function of some parameters. We
suppose that the first thing Eve does, just outside Alice's lab,
is a non-destructive measurement of the photon number. Sometimes,
she will simply find $n=0$ and there is nothing more to do. When
$n>0$, she will choose some attacks with the suitable
probabilities. We have attributed all the losses in the line to
Eve: this means that Eve replaces the quantum channel with a
lossless line, and takes advantage of the losses to keep in a
quantum memory or simply block some photons.

{\em Strategy for $n=1$}. When Eve finds one photon, with some
probability $p_{c1}$ she applies the well-known optimal incoherent
attack \cite{fuchs}, that consists in (i) applying the optimal
asymmetric phase-covariant cloning machine \cite{phasecov}, (ii)
forwarding the original photon to Bob while keeping the clone and
the ancilla in a quantum memory, (iii) make the suitable
measurement as soon as the basis is revealed. This strategy
contributes to Bob's detection rate with \ba
R_1&=&\demi\,\eta\,p_A(1)p_{c1}\,, \ea where the factor
$\demi\eta$ is due to the fact that Bob must accept the item
(detect the photon and accept at sifting). On these items, Eve
introduces a disturbance $D_1$ and gains the information
$I_1(D_1)=1-H(P_1)$ with $P_1=\demi+\sqrt{D_1(1-D1)}$. With
probability $p_{b1}=1-p_{c1}$, Eve simply blocks the photon
--- in principle, one can define the probability $p_{l1}$ that Eve
leaves the photon fly to Bob without doing anything, but this is
not useful for her (we left this parameter free in our numerical
simulations, see Section \ref{sec3}, and verified that one indeed
finds always $p_{l1}=0$).

{\em Strategy for $n=2$}. Sometimes, Eve finds two photons. The
standard PNS strategy is a storage attack: Eve keeps one photon in
a quantum memory, and forwards the other one to Bob. Eve applies
the storage attack with probability $p_{s2}$. This strategy
contributes to Bob's detection rate with \ba
R_{2s}&=&\demi\,\eta\,p_A(2)p_{s2}\,; \ea on these items, Eve
introduces no disturbance $D_1$ and gains the information
$I_{s2}=1$. As stressed in the introduction, the main theme of
this work is CL's observation that the storage attack may not
always be the best Eve can do on two photons. With probability
$p_{c2}$, she rather uses a $2\rightarrow 3$ asymmetric cloning
machine, keeps the clone and the ancillae and forwards the two
original photons, now slightly perturbed, to Bob. This strategy
contributes to Bob's detection rate with \ba
R_{2c}&=&\demi\,\big(1-(1-\eta)^2\big)\, p_A(2)p_{c2}\,; \ea on
these items, Eve introduces a disturbance $D_2$ and gains an
information $I_{2}(D_2)$ that depends on the cloning machine that
is used. Finally, one can in principle define the probability of
blocking both photons $p_{b2}$; but this turns out to be always
zero in practice (as for $p_{l1}$, we used this as a free
parameter in the numerical simulations). The reason is the
following. If Eve could reproduce Bob's detection rate by blocking
all the $n=1$ items (in which case, she might have to block also
some of the $n=2$ items), she'd have full information. Alice will
then choose her probabilities $p_A(n)$ in such a way that this is
not the case: Eve must be forced to forward some items with $n=1$.
Now, Eve gains more information on the $n=2$ than on the $n=1$
items: therefore, she has better use all the losses to block as
much $n=1$ items as possible; but then, she cannot block any $n=2$
item. Thus $p_{b2}=0$ and $p_{c2}=1-p_{s2}$.

{\em Strategy for $n\geq 3$}. If Eve finds more than two photons,
we suppose that she performs always the storage attack: she keeps
one photon and forwards the remaining $n-1$ photons to Bob. This
strategy contributes to Bob's detection rate with \ba
R_{3}&=&\demi\,\sum_{n\geq 3} \big(1-(1-\eta)^{n-1}\big)\,
p_A(n)\,; \ea on these items, Eve introduces no disturbance and
gains full information. This is not always optimal: unambiguous
discrimination strategies \cite{sarg,sarg2} or cloning attacks
\cite{cl} may give Eve more information. However, we don't discuss
the full optimization because in any case the contribution of
items where $n>2$ to the total information is small, as will be
clear below. Note also that in a storage attack Eve systematically
removes one photon; at very short distances, this might not be
possible because the expected losses in the line Alice-Bob are not
large enough. To avoid any surprise, we shall start all our
numerical optimization at a distance $d=10$km, where the losses
are definitely large enough to allow storage attack on all items
with $n\geq 3$ \cite{note3}.

{\em Summary.} We allow to perform different attacks with
different probabilities, conditioned on the knowledge of the
number of photons present in each pulse. Apart from the hypotheses
made on $R_3$, this represents the most general incoherent attack
on the BB84 protocol --- provided the hardware is protected
against "realistic attacks" like Trojan horse, faked states and
similar \cite{makarov}, as we suppose it to be.

\subsection{Formulae for Eve's attack}

We can now group everything together and describe the formulae
that will be used for Eve's attack. Eve's information on Bob's
bits reads \cite{note2} \ba I(A:E)&=& R_1I_1(D_1)+R_{2s}
+R_{2c}I_2(D_2)+R_3\label{ibe} \ea where \ba I_1(D_1)&=&1-
H\Big(\,\demi+\sqrt{D_1(1-D1)}\,\Big) \ea and where $I_2(D_2)$ is
the information gained by Eve using a $2\rightarrow 3$ asymmetric
cloning machine, for which the optimal is not known (see next
Section). For a given probability distribution used by Alice
$p_A(n)$, Eve chooses the four parameters $p_{c1}$, $p_{c2}$,
$D_1$ and $D_2$ in order to maximize (\ref{ibe}), submitted to the
constraints that determine $t$ and $V$. The {\em constraint on
$t$} guarantees that the losses introduced by Eve must be those
expected on the quantum channel, so in particular that Bob's
detection rate is unchanged: \ba R_1+R_{2s} +R_{2c}+R_3&=&
\demi\,\big(1-p_B(0)\big)\,.\label{constr1} \ea Alice and Bob have
to choose their source in order to ensure that Eve cannot set
$R_1=0$, otherwise she has full information by simply using the
PNS. This is the reason why the contribution of $R_3$ is small:
the leading term is a fraction of $p_A(1)$, typically of the order
of $p_A(2)$. Now, $p_{A}(3)/p_{A}(2)=O(\mu)\sim 0.1$ for the usual
Poissonian source, and even smaller for sub-Poissonian ones. The
{\em constraint on $V$} guarantees that the error rate introduced
by Eve must sum up to the observed optical QBER, that is \ba
R_1D_1+ R_{2c}D_2&=& \demi\,(1-p_B(0))\left(\frac{1-V}{2}
\right)\,.\label{constr2}\ea In the next Section, we discuss a
good choice of $I_2(D_2)$, then perform numerically the
optimization of Eve's strategies over the four parameters
$p_{c1}$, $p_{c2}$, $D_1$ and $D_2$. Before this, we are now able
to pinpoint the limitations of the analysis of CL.

\subsection{The limitation in CL}

In our notations, the parameter $p$ that characterizes Alice's
source in Ref. \cite{cl} is given by $p=\frac{p_A(1)}{1-p_A(0)}$,
the conditional probability of having one photon in a non-empty
pulse. Items with more than two photons are neglected, so in our
notations $R_3=0$ and $1-p=\frac{p_A(2)}{1-p_A(0)}$. This
assumption is not critical a priori. What is critical, is the
choice of Eve's attacks that are compared. The PNS attack
$R_{2c}=0$ is compared to a cloning attack in which not only
$R_{2s}$, but also $R_1$ is set to 0. As CL correctly note, the
comparison is fair only if the counting rates are the same between
the two strategies, which reads here $R_1^{PNS}+R_{2s}^{PNS}=
R_{2c}^{clon}$; in turn, this condition determines $p=1/(2-\eta)$.
Now, Alice should adapt the parameters of her probability
distribution as a function of the distance of the quantum channel.
Thus, a given value of $p$ will be optimal only for a given
distance (or at best, for a small range of possible distances):
the fact of setting $R_1=0$ in the cloning attack limits the
validity of CL's analysis to a given length of the line Alice-Bob.

In particular, if we consider that $p_A(n)$ is a Poissonian
distribution, then $\frac{1-p}{p}=
\frac{p(2|\mu)}{p(1|\mu)}=\frac{\mu}{2}$; setting $p=1/(2-\eta)$
leads to $\mu=\frac{2}{1-\eta}\geq 2$. This is a very large value
of $\mu$, that consequently can be used only at a very short
distance.

\section{Main results}
\label{sec3}

The problem that we want to solve involves a double optimization.
For any given distance, Alice should choose the parameters of her
source (e.g. for a Poissonian source, the mean number of photons
per pulse) in such a way as to optimize the secret key rate $S$,
Eq.~(\ref{rsk}). This quantity must be computed for Eve's best
strategy, i.e.~for $I(A:E)$ as large as possible: so, for any
choice of Alice's parameters, we must find the values of $p_{c1}$,
$p_{c2}$, $D_1$ and $D_2$ that maximize (\ref{ibe}) under the
constraints (\ref{constr1}) and (\ref{constr2}). For this task,
numerical algorithms are the reasonable choice. But, as an input
for these algorithms, we need the explicit form of $I_2(D_2)$. We
devote the next paragraph to this point.

\subsection{The choice of the $2\rightarrow 3$ cloning attack}

Eve receives two photons in the state $\ket{\psi}^{\otimes 2}$,
where $\ket{\psi}$ is one of the four states used in BB84. She has
these photons interact with a probe of hers, then she forwards
{\em two photons} to Bob, having introduced an average disturbance
$D_2$. By measuring her probe after the sifting phase, Eve gains
an information $I_2(D_2)$ on the state prepared by Alice. Finding
the optimal attacks means finding the best unitary transformation,
the best probe and the best measurement on it, such that
$I_2(D_2)$ is maximal for any given value of $D_2$. Though
well-defined, this problem is very hard to solve in general. Let's
restrict to attacks such that the photons flying to Bob after the
interaction are in a symmetric state, so that the transformation
reads \ba \ket{k} \ket{E}&\longrightarrow& \sum_{k'=1}^3c_{k}^{k'}
\ket{k'}\ket{E_k^{k'}} \ea where $\ket{k}$ is a basis of the
symmetric subspace of two qubits. There are nine vectors
$\ket{E_k^{k'}}$, so Eve's probe must be at least nine-dimensional
to avoid loss of generality. In addition, the measurement that
gives Eve the best guess on the state sent by Alice is not known
in general. In summary, finding the optimal $I_2(D_2)$ in full
generality amounts to solving an optimization over more than
hundred real parameters, for an undefined figure of merit. We give
this up and try a different approach, namely to guess a good (if
not the optimal) $2\rightarrow 3$ cloning attack.

Let's first look at what is already known. Two asymmetric
$2\rightarrow 3$ cloning machines were proposed in Ref.
\cite{sarg2}; Curty and L\"utkenhaus \cite{cl} based their
analysis of $2\rightarrow 3$ cloning attacks on those. The first
machine ({\em cloner A}) is a universal asymmetric cloner,
recently proven to be optimal in terms of fidelity \cite{sofyan}.
For a disturbance $D_2$ introduced on Bob's states, this machine
gives Eve an information \cite{cl} \ba
I_2^A(D_2)&=&2D_2+(1-2D_2)\big[1-H(P_2)\big] \ea with
$P_2=\demi\left(\sqrt{8D_2(1-4D_2)}\right)/(1-2D_2)$. A
particularly interesting feature is that $I_2^A(D_2=1/6)=1$. This
sounds at first astonishing, because one is used to Eve's getting
full information only by breaking all correlations between Alice
and Bob. But this is the case only if Eve receives a single photon
from Alice. Here Eve receives two photons in the same state. In
fact, the result $I_2^A(D_2=1/6)=1$ is not only reasonable, but it
can be reached by a much simpler strategy: Eve just keeps one of
the two incoming photons (so, after sifting, she can get full
information) and duplicates the second one using the optimal {\em
symmetric} $1\rightarrow 2$ cloner of Bu\v{z}ek-Hillery \cite{bh},
which makes copies with fidelity $\frac{5}{6}$, whence
$D_2=\frac{1}{6}$.

Cloner A is good (and we conjecture it to be optimal) to attack
two-photon pulses in the six-state protocol \cite{sixstate},
because of its symmetry. However, here we are dealing with BB84:
for the one-photon case, it is known that one can do better than
using the universal asymmetric cloner. In fact, the optimal
incoherent attack on single-photon pulses uses the {\em
phase-covariant} cloning machine, that copies at best two
maximally conjugated bases out of three \cite{phasecov}. So we
suspect that also for the $2\rightarrow 3$ cloning attack, we
should rather look for an asymmetric $2\rightarrow 3$
phase-covariant cloner. The second cloner ({\em cloner B})
described in Ref. \cite{sarg2} is an example of such a cloner.
However, it has some unpleasant features: one the one hand, in
terms of fidelity it is slightly suboptimal for the parameter that
defines symmetric cloning \cite{dariano}; more important,
$I_2^B(D_2)<1$ for all values of $D_2$ --- we don't write
$I_2^B(D_2)$ explicitly, because it is quite complicated and after
all unimportant for the present work; see Ref. \cite{cl}.

In summary, two $2\rightarrow 3$ asymmetric cloning machines have
been discussed in the literature, but they are suboptimal for our
task. Still, in the sake of comparison with Ref. \cite{cl}, we ran
our first numerical optimizations using $I_2^A(D_2)$, then
$I_2^B(D_2)$. The result is striking: (i) if $I_2=I_2^A$, then the
optimal strategy is always obtained for $D_2=\frac{1}{6}$,
whatever the values of the other parameters; (ii) if $I_2=I_2^B$,
the optimal strategy is the one that uses no $2\rightarrow 3$
cloning attack ($p_{c2}=0$). Following this observation, it is
natural to emit the following {\em conjecture}: the $2\rightarrow
3$ cloner is always used for the value of $D_2$ that gives \ba
I_2(D_2)&=&1\,.\label{eqconj}\ea Under this conjecture, we can
then replace $I_2(D_2)$ by 1 in (\ref{ibe}), and we have to find
the lowest value of $D_2$ for which (\ref{eqconj}) holds. In
general, this is a task of the same complexity as optimizing Eve's
strategy for all values of $D_2$; but we can at least construct a
very simple strategy which has an intuitive interpretation, and
which performs better than the ones which use cloners A and B:

{\em Hypothesis 4: the strategy for the $2\rightarrow 3$ cloning
attack is the following: out of two photons sent by Alice, Eve
keeps one and sends the other one into the optimal symmetric
$1\rightarrow 2$ phase-covariant cloner.}

This provides Eve with $I_{2}(D_2)=1$ after sifting, and Bob
receives two photons with a disturbance \ba D_2&= &\demi\left(1-
\frac{1}{\sqrt{2}}\right)\ea that is $\simeq 0.1464$
\cite{phasecov}. Since this disturbance is smaller than
$\frac{1}{6}$, for any fixed value of $V$ Eve can use the
$2\rightarrow 3$ cloning attack more often than in the optimized
version of the attack using cloner A, see constraint
(\ref{constr2}). That's why our new attack performs better.
Moreover, the attack has an intuitive form, that can be
generalized: in particular, it seems natural to extend the
conjecture to attacks on $n>2$ photons, although here we don't
consider this extension because these cases are rare (see above).
In what follows, we comment on the explicit results that we find
for the numerical optimization using this strategy.

\begin{center}
\begin{figure}
\epsfxsize=5cm \epsfbox{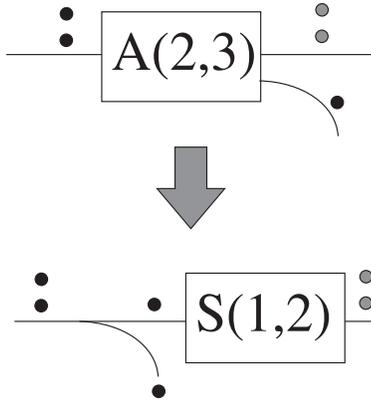} \caption{Illustrating the
conjecture on Eve's $2\rightarrow 3$ cloning strategies: the
asymmetric $2\rightarrow 3$ cloning machine $A(2,3)$ is actually
used at a working point where Eve keeps a perfect copy and
forwards two identically perturbed photons to Bob, produced with
the symmetric $1\rightarrow 2$ cloning machine $S(1,2)$.}
\label{figc23}
\end{figure}
\end{center}

\subsection{Numerical optimization for Poissonian sources}
\label{ssnum}

We use numerical optimization to find, under Hypotheses 1-4, Eve's
best strategy and the optimal value of Alice's parameters. We
consider a Poissonian distribution for Alice's source, \ba
p_A(n)\,\equiv\,p(n|\mu) &=&e^{-\mu}\,\frac{\mu^n}{n!}
\label{poiss}\ea so that the only parameter that characterizes
Alice's source is the mean number of photons $\mu$ (see
\ref{ssubpoix} below for extension to sub-Poissonian sources). As
sketched above, the numerical optimization is done as follows. For
any value of the distance $d$ Alice-Bob, we choose a value of
$\mu$ and find the values of $p_{c1}$, $p_{c2}$ and $D_1$ that
optimize Eve's information under the constraints. This gives a
value for the secret key rate $S$. Then we vary $\mu$ and repeat
the procedure, until the highest value of $S$ is found. This
defines the optimal value of $\mu$.

We have done these calculations for the nowadays standard (and
even conservative) values $\alpha=0.25$ dB/km, $\eta=0.1$ and
$p_d=10^{-5}$. Of course, the qualitative features are independent
of these precise values.

\begin{center}
\begin{figure}
\epsfxsize=8cm \epsfbox{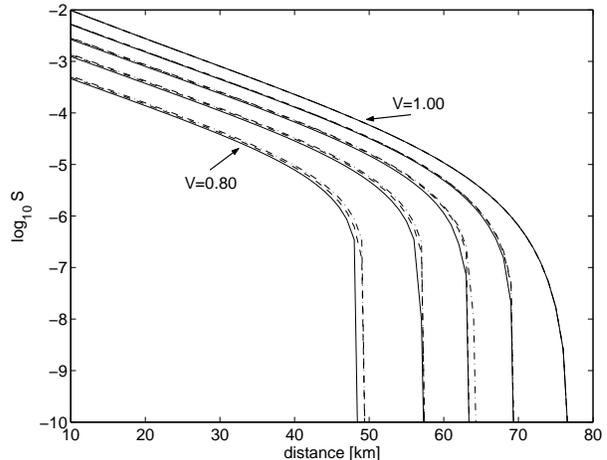} \caption{Secret key rate
per pulse $S$ as a function of the distance, for $\alpha=0.25$
dB/km, $\eta=0.1$ and $p_d=10^{-5}$, and for
$V=1,0.95,0.9,0.85,0.8$. The best attack (full line) uses Strategy
C for $2\rightarrow 3$ cloning; the value of the optimal $\mu$ is
fixed by this strategy. For comparison, we plot the results that
one would obtain using Strategy A for $2\rightarrow 3$ cloning
(dashed lines) and without using any $2\rightarrow 3$ cloning
(dashed-dotted lines), computed for the same $\mu$.}
\label{figexact}
\end{figure}
\end{center}

The achievable secret key rate $S$, Eq.~(\ref{rsk}), is plotted in
Fig.~\ref{figexact} as a function of the distance, in log scale.
The full lines are obtained by allowing Eve to use our new
$2\rightarrow 3$ cloning attack defined above. Supposing this
attack we can extract, at any distance, an optimal value of $\mu$:
this is the mean number of photons Alice and Bob should choose.
For the so-computed $\mu$, we then compute $S$ by supposing two
suboptimal attacks by Eve, namely no $2\rightarrow 3$ cloning, and
$2\rightarrow 3$ cloning with cloner A \cite{cl}. The results of
these suboptimal attacks are plotted in the discontinuous lines.
We see that indeed our strategy yields the best results for Eve
(the smallest $S$ achievable), but the difference between the
optimal and the suboptimal attacks is very small --- in fact,
under the assumptions of practical cryptography this difference is
completely negligible, see beginning of Section \ref{sec4}.

\begin{center}
\begin{figure}
\epsfxsize=8cm \epsfbox{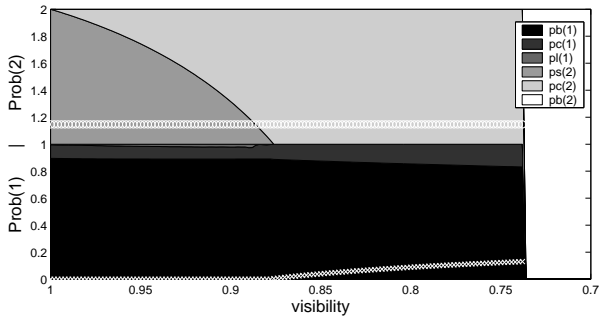} \caption{Probabilities that
define Eve's optimal attack as a function of the visibility, for
$d=30$km, for the optimal $\mu$. In the lower half one reads the
probabilities for the attacks on $n=1$; in the upper half, for
$n=2$. The white symbols represent $D_1$ (lower half) and $D_2$
(upper half). Note that high visibility (small optical errors) are
on the left. See text for detailed comment.} \label{figprob30km}
\end{figure}
\end{center}

\begin{center}
\begin{figure}
\epsfxsize=8cm \epsfbox{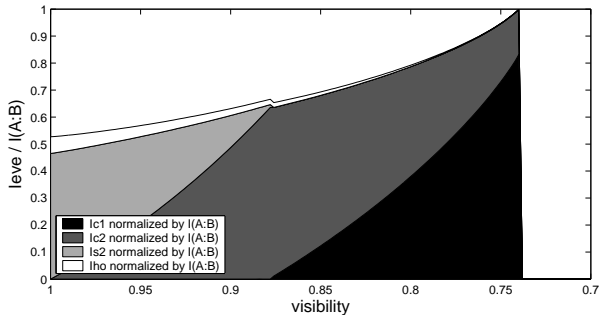} \caption{The four terms
that sum up to Eve's information (\ref{ibe}) as a function of the
visibility, for $d=30$km and for the optimal $\mu$. Eve's
information is divided by the value of $I(A:B)$ at any $V$. See
text for detailed comments.} \label{figieve30km}
\end{figure}
\end{center}

Figures \ref{figprob30km} and \ref{figieve30km} illustrate in
detail the parameters for Eve's optimal attack, for a fixed
distance (30 km), as a function of the visibility $V$. In
Fig.~\ref{figprob30km} are plotted the probabilities introduced in
paragraph \ref{sseve} that define Eve's strategies on the pulses
with $n=1$ (lower half of the figure) and with $n=2$ (upper half).
Fig.~\ref{figieve30km} represents the four terms that sum up to
Eve's information (\ref{ibe}). Much information is stored in these
graphics:

\begin{itemize}
\item First note that at $V\lesssim 0.74$, that is $Q_{opt}\gtrsim
13\%$, one has $I(A:E)=I(A:B)$ so $S=0$. For smaller values of the
visibility, with our assumptions on the attacks and on the
numerical values of the parameters, the BB84 protocol becomes
insecure for all $\mu$ at 30km. This is due to the characteristics
of the source: recall that for incoherent attacks on the BB84
protocol with perfect single-photon sources, the critical
visibility is $V\lesssim 0.7$ ($Q_{opt}\geq 14.67\%$) independent
of the distance \cite{review,fuchs}.

\item For $V=1$, Eve is not allowed to introduce any error.
Therefore, for $n=1$ she can either block or forward the pulse
without introducing any error ($D_1=0$), and she gains no
information; for $n=2$, she can only perform the storage attack.

\item As soon as $V<1$, Eve's strategy on the one-photon pulses
does not change, while on the two-photon pulses she starts using
the cloning strategy. She uses it on as many pulses as possible,
compatible with constraint (\ref{constr2}). This situation goes on
until $V\simeq 0.88$: for that visibility, Eve can perform the
$2\rightarrow 3$ cloning attack on all the two-photon pulses.
Then, for $V\lesssim 0.88$, Eve can start introducing errors (and
gaining some information) on one-photon pulses as well; and
indeed, we see the increase of $D_1$ in Fig.~\ref{figprob30km} and
the corresponding increase of $I_{c1}$ in Fig.~\ref{figieve30km}.

\item In the region $0.88\lesssim V<1$, we note an ambiguity of
the simulation for the single-photon pulses. In fact
(Fig.~\ref{figprob30km}) we have $p_{c1}>0$ but $D_1=0$, so this
"cloning" actually amounts to leaving photons undisturbed and
might as well be accounted for through $p_{l1}$. Recall that in
paragraph \ref{sseve} we said that one can always set $p_{l1}=0$;
it is now clear why: as long as $D_1=0$, letting pass is
equivalent to cloning; and we see that when $D_1$ becomes larger
than 0, cloning is applied on all the forwarded photons so that
indeed $p_{l1}=0$.

\item There is a slight discontinuity in Eve's information,
visible in Fig.~\ref{figieve30km}, at the point where Eve starts
to use the cloning strategy on the single-photon pulses. We ran
more detailed simulations in order to rule out the possibility
that this is an artefact. It appears that this discontinuity is a
direct consequence of a discontinuous modification of $\mu$: for
that value of the parameters, Alice and Bob should decrease $\mu$
slightly more than expected by continuity.

\end{itemize}

At the end of this discussion, one might reasonably raise a doubt.
We have just seen that the $2\rightarrow 3$ cloning machine is
used as soon as $V<1$, and that for some rather high visibility
($V\approx 0.88$ at $d=30$km) it is used on all the two-photon
pulses. Why then is its effect so negligible in comparison to the
case when this machine is not used, as we saw in
Fig.~\ref{figexact}? The reason is that Figs \ref{figprob30km} and
\ref{figieve30km} would look fundamentally different if the
$2\rightarrow 3$ cloning machine is not used. If Eve performs the
storage attack instead of the cloning attack on the two-photon
pulses, then she can introduce errors, and consequently gain
information, on the single-photon pulses: we'd have $D_1>0$ and
$I_{c1}>0$ as soon as $V<1$, not only for $V\lesssim 0.88$. It
turns out that all the information, that Eve loses on the
two-photon pulses by not using the cloning attack, is almost
exactly compensated by the information that she gains on the
single-photon pulses. This casts a new light on the result of
Fig.~\ref{figexact}: the difference between the optimal and the
suboptimal strategies is small, not because the $2\rightarrow 3$
cloning is rarely used, but because the constraints
(\ref{constr1}) and (\ref{constr2}) imply that using the
$2\rightarrow 3$ cloning attack on $n=2$ reduces the possibility
of using the $1\rightarrow 2$ cloning attack on $n=1$.

\section{Extensions and remarks}
\label{secnew}

\subsection{Extension to sub-Poissonian sources}
\label{ssubpoix}

For the numerical optimization, we have supposed the Poissonian
distribution for the number of photons produced by Alice, because
this is the most frequent case in practical implementations.
However, sub-Poissonian sources are being developed for quantum
cryptography \cite{single}. The main result, namely that
$2\rightarrow 3$ cloning attacks contribute with a very small
correction to Eve's information, remains valid for these sources:
the fraction of pulses with $n=2$ photon is even smaller than in
the Poissonian case, so the contribution of the $2\rightarrow 3$
cloning attack will be even more negligible --- actually, it is
even possible that, for a sufficiently large deviation from the
Poissonian behavior, this kind of attack does not help at all.

\subsection{Extension to other protocols}

One might ask how our study applies to other protocols. In the
last months, practical QKD has witnessed great progress: several
ideas have been put forward that make the PNS attacks less
effective by modifying the hardware \cite{decoy}, the classical
encoding \cite{sarg,sarg2} or the quantum encoding \cite{moreph}.
Of course, even if the PNS can never be used by Eve, multi-photon
pulses open the possibility for elaborated cloning attacks: these
must be taken into account when assessing the security of new
protocols.

\subsection{About reverse reconciliation}
\label{ssieb}

In Section \ref{sec2}, when defining $S$ in (\ref{rsk}), we
mentioned the fact that $I(B:E)$ is slightly smaller than $I(A:E)$
here, so that Alice and Bob would better do "reverse
reconciliation" \cite{reverse}. In this paragraph, we want to
elaborate a little more on this point.

The first cause of the relation $I(B:E)<I(A:E)$ is the presence of
dark counts: when Bob accepts an item, Eve (as well as Bob
himself) does not know if his detector fired because of the photon
that she has forwarded (and on which she has some information) or
because of a dark count (on which she has no information). It is
easy to take this effect into account. Suppose that Eve forwards
$n$ photons to Bob. Conditioned to this knowledge, Bob's detection
rate reads $r_n=r_{ph}+r_{dark}$ where $r_{ph}=(1-(1-\eta)^n)$ and
$r_{dark}=p_d\,(1-\eta)^n$. Thus, to obtain $I(B:E)$, the
$n$-photon contribution to formula (\ref{ibe}) should be
multiplied by a factor $(1-H(\epsilon_n))$, where
$\epsilon_n=r_{dark}/r_n$. Now, $\epsilon_1\simeq p_d/\eta$, and
$\epsilon_{n\geq 2}<\epsilon_1$; so all these corrections are
really negligible.

The second contribution is much less easily estimated: it comes
from the $2\rightarrow 3$ cloning machines. The formulae we used
for Strategies A and B, derived by CL \cite{cl}, refer to the
mutual information Alice-Eve. In Strategy C, that looks optimal
when $I(A:E)$ is optimized, Eve's information on Bob's result is
smaller than 1 because she does not know deterministically whether
Bob will obtain the same bit as Alice or the wrong bit. This study
would require some more work. We don't think this work is worth
while doing, after seeing how small is the correction introduced
on the final values of $\mu$ and $S$ by taking the $2\rightarrow
3$ cloning attack into account.

\section{Analytical formulae for rapid estimates}
\label{sec4}

\subsection{Further simplifying assumptions}

As mentioned before, the goal of this Section is to provide some
simple formulae that allow a good estimate of the important
parameters (optimal mean number of photons, expected secret key
rate $S$, maximum distance) for implementations of the BB84
protocol, without resorting to the full numerical optimization.
Indeed, for practical implementations, absolute precision of these
calculations is not required: on the one hand, existing algorithms
for error correction and privacy amplification (EC+PA) reach up to
some 80\% of the attainable $S$; on the other hand, nobody is
going to operate his crypto-system too close to the critical
distance. So in short, what one needs is (i) an estimate of the
critical distance in order to keep away from it, (ii) an estimate
of the optimal mean number of photons per pulse in order to
calibrate the source, and (iii) an estimate of the secret-key rate
(of Eve's information) in order to choose the parameters for
EC+PA. Note that similar formulae have been found by L\"utkenhaus
\cite{lutkenhaus}; in that work, however, Eve was supposed to have
an influence on {\em all} the sources of inefficiency, in
particular the parameters of the detector. This is why we can't
simply refer to L\"utkenhaus' results here.

Thus, for this analysis, we make two further simplifying
assumptions on Eve's attack, namely:
\begin{enumerate}
\item We neglect completely the contribution of the pulses with
$n\geq 3$ photons. Since we are interested in sources where the
mean number of photons $\mu$ is significantly smaller than 1, we
have \ba p_A(1)=\mu&\;,\;& p_A(2)=g_2\,\frac{\mu^2}{2}\ea whence
in particular $1-p_B(0)\approx \mu t\eta$. The factor $g_2$ is 1
for a Poissonian source, smaller than 1 for sub-Poissonian
sources.

\item For $n=2$, we neglect the $2\rightarrow 3$ cloning attack
and focus only on storage attacks, that is $p_{2s}=1$. In fact, we
have seen that the cloning attack plays a non-negligible role only
for $V\approx 0.8$; but this means an optical QBER of 10\%, which
is enormous and would lead to the failure of the EC+PA algorithms.
For practical cryptography, $V\gtrsim 0.9$ is required, and in
this region the correction due to cloning $2\rightarrow 3$ is
really negligible.
\end{enumerate}

\subsection{$S$ as a function of $\mu$ alone}
\label{ssratemu}

Using the Poissonian distribution (\ref{poiss}), the mutual
information Alice-Bob (\ref{iab}) reads \ba
I(A:B)&=&\demi\,\big(\mu t\eta
+2p_d\big)\,\big(1-H(Q)\big)\label{iab2} \ea with the QBER \ba
Q&=& \demi- \frac{V}{2\left(1+\frac{2p_d}{\mu
t\eta}\right)}\,\label{qexpl}. \label{qber2}\ea Using our
assumptions $R_{2c}=R_3=0$, the first constraint (\ref{constr1})
that Eve must fulfill reads $\mu\,
p_{c1}\,\eta+g_2\frac{\mu^2}{2}\eta=\mu t\eta$, whence one can
extract \ba p_{c1}&=&t\,-\,g_2\frac{\mu}{2}\,. \ea The second
constraint (\ref{constr2}), using $R_{2c}=0$ and the expression we
have just found for $p_{c1}$, reads
$\mu\left(t-g_2\frac{\mu}{2}\right)\eta D_1\,=\, \mu t\eta
\left(\frac{1-V}{2}\right)$, whence \ba D_1&=&
\frac{1-V}{2-g_2\mu/t}\,. \ea Then, the mutual information
Alice-Eve (\ref{ibe}) reads \ba I(A:E)&=& \demi\,\mu\eta\,
\left[\left(t-g_2\frac{\mu}{2}\right)I_1(D_1)+g_2\frac{\mu}{2}\right]
\label{ibe2} \ea where we recall that $I_1(D_1)=1-H(P_1)$ with
$P_1=\demi+\sqrt{D_1(1-D1)}$.

Presently then, $S=I(A:B)-I(A:E)$ is written as a function
$S(\mu)$ of $\mu$ alone --- in particular, our hypotheses removed
two of the four parameters of Eve's attacks, and because of the
two constraints there are no more free parameters for Eve. One can
then find the optimal $\mu$ as a function of the distance, and the
corresponding $S$, by running a numerical optimization of
$S(\mu)$. This is already simple enough and gives very accurate
results, see Fig. \ref{figapprox}. Still, we want to go a few
steps forward, to provide less accurate but explicit formulae.

\subsection{Formulae for high visibility and not too long distances}

To perform analytical optimization, we must get rid of the
$\mu$-dependence in the non-algebraic functions $H(Q)$ and
$I_1(D_1)$. This can be done for not too long distances, that is
when $2p_d\ll\mu t\eta$, because then $Q\simeq
Q_{opt}=\frac{1-V}{2}$. Moreover, one can easily see that for
$V=1$, the optimal $\mu$ (satisfying $\frac{dS}{d\mu}=0$) is \ba
\mu\,=\,\frac{t}{g_2}&\;\;&\mbox{  $(V=1)$}\,. \ea Therefore, we
set this value for $\mu$ in $D_1$, so that now $D_1=1-V=2Q_{opt}$
becomes also independent of $\mu$ \cite{note4}. This gives
$P_1\equiv P= \demi-\sqrt{V(1-V)}$. Under these new assumptions
\ba S(\mu)&\simeq &
\demi\,\mu\eta\,\left[t\big(H(P)-H(Q_{opt})\big)-
g_2\frac{\mu}{2}H(P)\right]\,; \label{sexpl1}\ea the maximum is
obtained for $\frac{dS}{d\mu}=0$, that is \ba \mu& \simeq &
\frac{t}{g_2}\,\left(1-\frac{H(Q_{opt})}{H(P)}\right)\,.
\label{muexpl}\ea This must be non-negative, so this approximation
(in particular here, the approximation $\mu=t/g_2$ in $D_1$) is
valid provided $Q_{opt}<P$, that is for $V>0.8$; as we discussed
in the introduction of this Section, this is perfectly consistent
with the visibility requirements in practical setups. Inserting
(\ref{muexpl}) into (\ref{sexpl1}), we find an explicit formula
for the secret key rate \ba S &\simeq &\frac{1}{4}\,\eta\,
\frac{t^2}{g_2} \,H(P)\,
\left(1-\frac{H(Q_{opt})}{H(P)}\right)^2\,.\label{sexpl} \ea In
the limiting case $V=1-\varepsilon$ we can set $H(P)=1$ while
$H(Q_{opt})=H(\varepsilon/2)$ cannot be neglected because $H$
increases very rapidly for its argument close to zero. Therefore
\ba S &\simeq &\frac{1}{2}\,\eta\,t\, \frac{t}{g_2}
\left(\demi-H(\varepsilon/2)\right)\quad (V=1-\varepsilon)\,.\ea
This formula has an intuitive meaning \cite{note0}:
$\frac{1}{2}\,\eta\,t\, \frac{t}{g_2}$ is simply the sifted-key
rate; $H(\varepsilon/2)$ is the fraction that must be subtracted
in error correction, and a fraction $\demi$ is subtracted in
privacy amplification because of the PNS attack \cite{note4}.

For distances far from the critical distance, the agreement of
both (\ref{muexpl}) and (\ref{sexpl}) with the exact results is
again satisfactory (Fig. \ref{figapprox}).

\subsection{Exact limiting distance for $V=1$}

\begin{center}
\begin{figure}
\includegraphics[width=8cm]{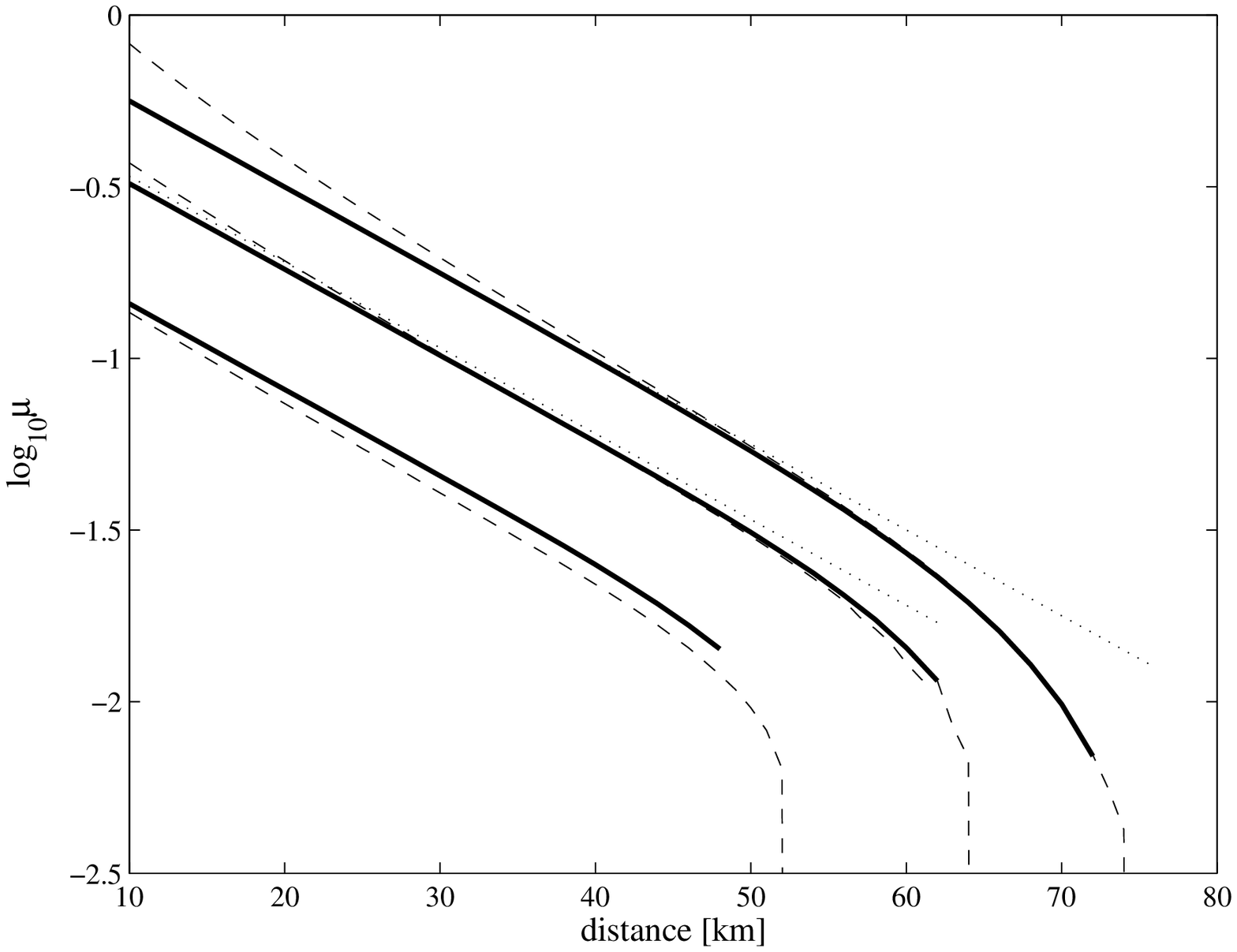}
\vspace{2mm}
\includegraphics[width=8cm]{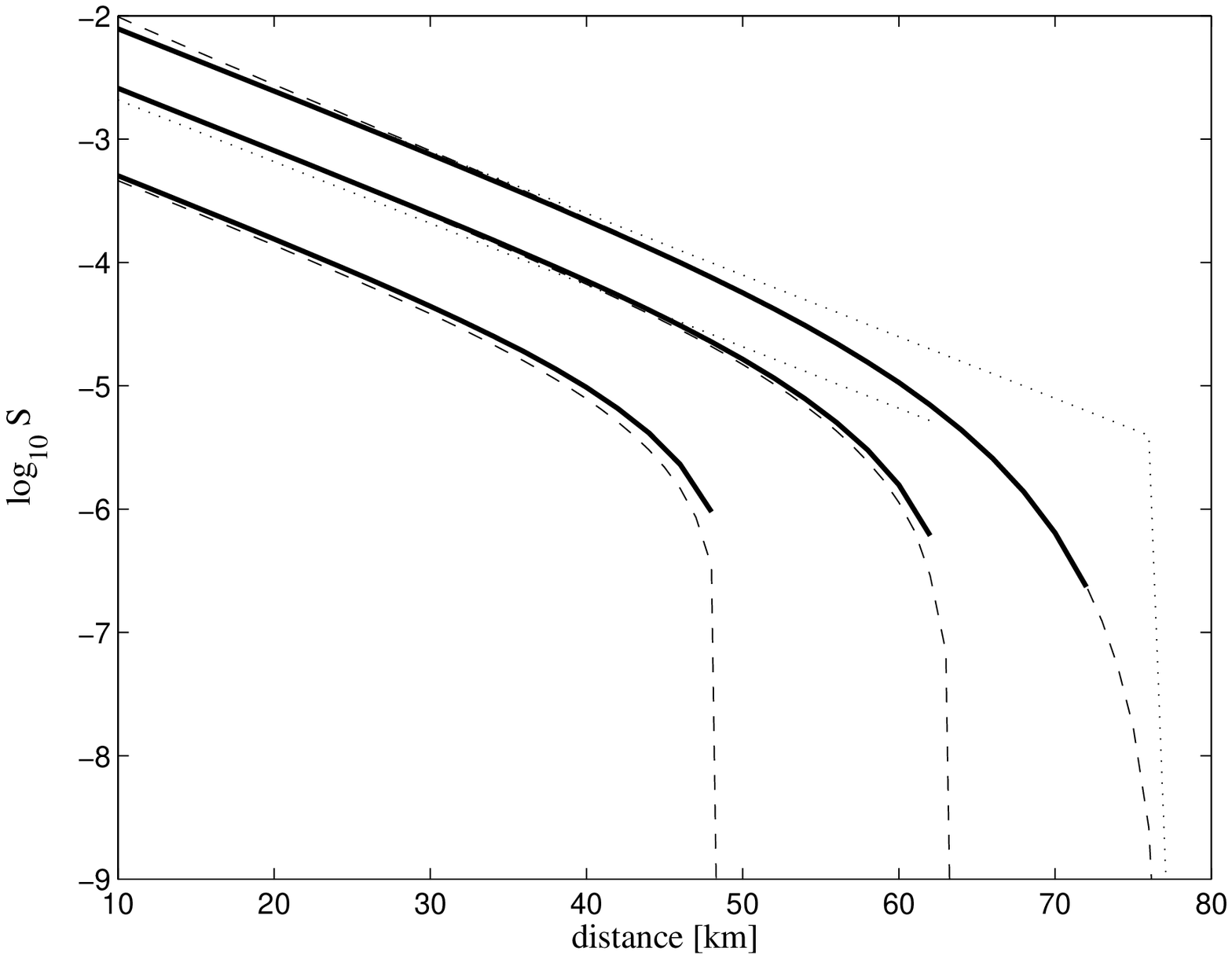}
\caption{Optimal $\mu$ and secret key rate per pulse $S$ (log
scale) for Poissonian sources as a function of the distance, for
$\alpha=0.25$, $\eta=0.1$ and $p_d=10^{-5}$, and for
$V=1,0.9,0.8$. Comparison of the exact results (dashed lines,
coming from Fig. \ref{figexact}) with two approximations: (I) Full
lines: numerical optimization over $\mu$ alone as discussed in
paragraph \ref{ssratemu}. (II) Dotted lines: explicit formulae
(\ref{muexpl}) and (\ref{sexpl}), that cannot be used for $V=0.8$.
For $V=1$, the vertical asymptote is the limiting distance defined
by (\ref{tc}).} \label{figapprox}
\end{figure}
\end{center}

For the value of the limiting distance, we were able to find a
closed formula only for the case $V=1$. The idea is that $\mu$
decreases very rapidly when approaching the limiting distance, so
that now $\mu t\eta \ll p_d$. The QBER (\ref{qber2}) becomes
$Q=\demi-\varepsilon$ with $\varepsilon=\frac{\mu t\eta}{4p_d}$
\cite{noteqber}. Now, it holds
$1-H\left(\demi-\varepsilon\right)=\frac{2}{\ln
2}\varepsilon^2+O(\varepsilon^4)$. Inserting this into
(\ref{iab2}) we obtain \ba I(A:B)&=&p_d\,\frac{1}{8 \ln
2}\left(\frac{\mu t\eta}{p_d}\right)^2+O(\mu t\eta)^3\,. \ea On
the other hand, $I(A:E)$ is still given by (\ref{ibe2}), of course
with $I_1(D_1)=0$ since $V=1$, so
$I(A:E)=\frac{1}{4}g_2\eta\mu^2$. The limiting distance is thus
defined by imposing $I(A:B)=I(A:E)$ i.e. $S=0$, that is, by the
attenuation \ba t_{lim} &=& \sqrt{2\ln
2\,g_2\,\frac{p_d}{\eta}}\,. \label{tc}\ea This result is in good
agreement with the limiting distance found in the exact
calculation, see Fig. \ref{figapprox}. The calculation of
(\ref{tc}) is easy because $\mu$ drops out of the condition $S=0$;
this is no longer the case for $V<1$, that's why the estimate of
the limiting distance becomes cumbersome: one has to provide the
link between $\mu$ and $t$ when approaching that distance,
different from (\ref{muexpl}).

\section{Conclusion}
\label{sec5}

In conclusion, we have discussed incoherent attacks on the BB84
protocol in the presence of multi-photon pulses that allow both
for the photon-number splitting and the $2\rightarrow 3$ cloning
attacks. We have identified a new efficient $2\rightarrow 3$
cloning attack: Eve keeps one of the incoming photons, and sends
the other one into the suitable symmetric $1\rightarrow 2$ cloner,
then forwards the two photons to Bob. The effect of taking the
cloning attacks into account is negligible for realistic values of
the parameters (in particular, for an optical visibility $V\gtrsim
0.9$) with respect to the PNS attacks. This means that these
attacks do not change the security of BB84; however, they may be
important when assessing the security of modified protocols aimed
at countering the PNS attacks.

\section{Acknowledgements}

A.N. acknowledges the hospitality of the Group of Applied Physics
of the University of Geneva, where this work was done, and is
grateful to Marc-Andr\'e Dupertuis (EPFL) for acting as internal
reporter for his master thesis. This work was supported by the
Swiss NCCR "Quantum photonics" and the European Union Project
SECOQC.

\end{multicols}


\begin{thebibliography}{99}

\bibitem{review} N. Gisin, G. Ribordy, W. Tittel, H. Zbinden, Rev. Mod. Phys {\bf 74}, 145
(2002)

\bibitem{brassard} G. Brassard, N. L\"{u}tkenhaus, T. Mor, B.C. Sanders, Phys. Rev. Lett. {\bf 85}, 1330
(2000)

\bibitem{lutkenhaus} N. L\"{u}tkenhaus, Phys. Rev. A {\bf 61}, 052304 (2000)

\bibitem{bb84} C.H. Bennett, G. Brassard, in: {\em
Proceedings IEEE Int. Conf. on Computers, Systems and Signal
Processing, Bangalore, India} (IEEE, New York, 1984), pp. 175-179.


\bibitem{cl} M. Curty, N. L\"utkenhaus, Phys. Rev. A {\bf 69}, 042321 (2004)

\bibitem{sarg} V. Scarani, A. Ac\'{\i}n, G. Ribordy, N. Gisin, Phys. Rev. Lett. {\bf 92}, 057901
(2004)

\bibitem{sarg2} A. Ac\'{\i}n, N. Gisin, V. Scarani, Phys. Rev. A {\bf 69},
012309 (2004)


\bibitem{decoy} Recently, a number of new
protocols have been invented using the notion of "decoy states",
and which are secure against the PNS attacks: W.-Y. Hwang, Phys.
Rev. Lett. {\bf 91}, 057901; H.-K. Lo, X. Ma, K. Chen,
quant-ph/0411004; X.-B. Wang, quant-ph/0410075 and
quant-ph/0411047.


\bibitem{single} A. Beveratos, R. Brouri, T. Gacoin,
A. Villing, J.-P. Poizat, P. Grangier, Phys. Rev. Lett. {\bf 89},
187901 (2002); E. Waks, K. Inoue, C. Santori, D. Fattal, J.
Vuckovic, G. S. Solomon, Y. Yamamoto, Nature {\bf 420}, 762
(2002); R. All\'eaume et al., New J. Phys. {\bf 6}, 92 (2004)


\bibitem{herald} O. Alibart, S. Tanzilli, D.B. Ostrowsky, P.
Baldi, quant-ph/0405075; S. Fasel et al., New J. Phys. {\bf 6},
163 (2004)


\bibitem{note5} This formula is rigorously correct if $1-p_B(0)\approx
p_B(1)$; in fact, if Bob receives two photons, the effect of $V<1$
is to increase slightly the possibility of double count. For the
distances that we are going to consider, $p_B(1)\gg p_B(2)$ indeed
holds.

\bibitem{note6} Recall the intuitive argument given in the
introduction: if Eve could set $\eta=1$ when it is convenient for
her, there would be no advantage for her in sending out two
photons instead of one, because Bob detects the itme anyway.

\bibitem{singap} We consider only one-way communication protocols
for error correction and privacy amplification. For two-way
protocols ("advantage distillation") an avantageous coherent
strategy has been found [D. Kaszlikowski, J.Y. Lim, L.C. Kwek,
B.-G. Englert, quant-ph/0312172], but it has been proved recently
that the same result can be achieved by individual attacks and
measurements provided Eve waits until the end of the advantage
distillation procedure [A. Ac\'{\i}n et al., quant-ph/0411092].




\bibitem{uncond} For the most advanced lower bounds
on BB84 with practical devices, see: D. Gottesman, H.-K. Lo, N.
L\"utkenhaus, J. Preskill, Quant. Inf. Comput. {\bf 4}, 325
(2004). Note however that their bounds cannot be compared directly
to the ones described in this paper, since they suppose that Eve
has full control over the imperfections of Bob's detectors.

\bibitem{real} S. F\'elix, N. Gisin, A. Stefanov, H. Zbinden, J. Mod. Opt. {\bf
48}, 2009 (2001); M. Williamson, V. Vedral, J. Mod. Opt. {\bf 50},
1989 (2003); M. Curty, N. L\"utkenhaus, quant-ph/0411041

\bibitem{csi} I. Csisz\'ar and J. K\"{o}rner, IEEE Trans. Inf. Theory
{\bf IT-24}, 339 (1978).

\bibitem{fuchs} C.A. Fuchs, N. Gisin, R.B. Griffiths, C.-S. Niu, A.
Peres, Phys. Rev. A {\bf 56}, 1163 (1997)

\bibitem{phasecov} R.B. Griffiths, C.-S. Niu, Phys. Rev. A {\bf 56}, 1173 (1997) D.
Bru\ss, M. Cinchetti, G.M. D'Ariano, C. Macchiavello, Phys. Rev. A
{\bf 62}, 012302 (2000).

\bibitem{note3} This distance of 10km is a conservative distance,
the details depend on the probabilities $p_A(n)$. For instance,
let's consider Poissonian distribution, and the even stronger
condition that Eve can always keep one photon whenever $n>1$. This
is possible as soon as $p(1|\mu)\eta
+\sum_{n>1}p(n|\mu)(1-(1-\eta)^{n-1})$, which is Bob's rate when
Eve forwards all the items $n=1$ and keeps always a photon for
$n>1$, becomes equal to $1-p(0|\mu t\eta)$, the expected Bob's
rate in the absence of Eve. For $\eta\geq 0.1$, the condition
holds for any value of $\mu$ as soon as $t\lesssim 0.7$, that
means for $d\gtrsim 6$km for $\alpha=0.25$ dB/km.

\bibitem{note2} We neglect a small contribution to $I(A:E)$. In
fact, Eve can gain some information on Alice's bit even from the
single-photon pulses that she does not forward to Bob, if Bob has
a dark count and accepts the item. This contribution is completely
negligible, because it is of the order $\sim p_dp_A(1)$, compared
to $R_{c1}\sim \eta p_A(1)$.

\bibitem{makarov} A. Vakhitov, V. Makarov, D. R. Hjelme, J. Mod. Opt. {\bf 48}, 2023
(2001); V. Makarov, D. R. Hjelme, J. Mod. Opt. (to be published,
2004).


\bibitem{sofyan} S. Iblisdir, A. Ac\'\i n, N. Gisin, J.
Fiur\'a\v{s}ek, R. Filip, N. J. Cerf, quant-ph/0411179

\bibitem{bh} V. Bu\v{z}ek, M. Hillery, Phys. Rev. A {\bf 54}, 1844 (1996)

\bibitem{sixstate} D. Bru\ss, Phys. Rev. Lett. {\bf 81}, 3018
(1998); H. Bechmann-Pasquinucci, N. Gisin, Phys. Rev. A {\bf 59},
4238 (1999)


\bibitem{dariano} G.M. D'Ariano, C. Macchiavello, Phys. Rev. A {\bf 67}, 042306 (2003)

\bibitem{moreph} Z.D. Walton, A.F. Abouraddy, A.V. Sergienko, B.E.A. Saleh, M.C.
Teich, Phys. Rev. Lett. {\bf 91}, 087901 (2003); J.C. Boileau, D.
Gottesman, R. Laflamme, D. Poulin, R.W. Spekkens, Phys. Rev. Lett.
{\bf 92}, 017901 (2004); X.-B. Wang, quant-ph/0406100

\bibitem{reverse} F. Grosshans, N.J. Cerf, J. Wenger, R. Tualle-Brouri, P. Grangier,
Quant. Inf. Comput. {\bf 3}, 535 (2003); D. Collins, N. Gisin, H.
de Riedmatten, quant-ph/0311101

\bibitem{note4} Note that the
relation $D_1=2Q$ under these assumptions is perfectly consistent:
indeed, if $\mu=t/g_2$, then $p_{c1}=t/2$: Eve introduces errors
in half of the transmitted photons. Consequently, she can
introduce a double disturbance on these items.

\bibitem{note0} We thank N. L\"utkenhaus for bringing this point
to our attention.

\bibitem{noteqber} At first sight, the condition $Q\approx \demi$
seems at odds with the well-known bound of $Q=14.67\%$ for
incoherent attacks, beyond which the key distribution becomes
insecure \cite{review,fuchs}. However, there is no contradiction:
the bound concerns the optical QBER, that in our case is zero
($V=1$). The error rate due to dark counts may become larger than
$14.67\%$, since these errors are not useful for Eve.






\end{thebibliography}
\end{document}